\documentclass[aps,prl,reprint,groupedaddress]{revtex4-2}
\usepackage{xr}

\usepackage{amsmath}
\usepackage{amssymb}
\usepackage{braket}
\usepackage{bbm}
\usepackage{graphicx}
\usepackage{dcolumn}
\usepackage{bm}
\usepackage{tikz}
\usepackage{placeins}
\usepackage{nicefrac}
\usepackage{makecell}
\usepackage{caption}
\usepackage{subcaption}
\usepackage{verbatim}
\usepackage[normalem]{ulem}
\usepackage{xurl}
\usepackage[dvipsnames]{xcolor}
\usepackage{array}
\usepackage{multirow}
\usepackage{booktabs}
\usepackage{comment}
\usepackage{amsthm}
\usepackage{mwe}
\usepackage{mathrsfs}
\allowdisplaybreaks

\newcommand{\refA}[1]{#1}
\newcommand{\refB}[1]{#1}
\newcommand{\refC}[1]{#1}
\newcommand{\refD}[1]{#1}

\usepackage{hyperref}
\hypersetup{
    colorlinks=true,
    linkcolor=blue,
    filecolor=blue,      
    urlcolor=blue,
    citecolor = blue,
    breaklinks=true,
}

\newcommand{\kB}{k_{\text{B}}}

\begin{document}

\title{Universal Dynamical Response to Slow Driving in Chaotic Systems}
\author{Nachiket Karve}
\email{nachiket@bu.edu}
\author{Nathan Rose}
\author{David Campbell}
\author{Anatoli Polkovnikov}
\affiliation{Department of Physics, Boston University, Boston, Massachusetts 02215, USA}
\date{\today}

\begin{abstract}
    We propose a unified perspective on classical and quantum chaos based on the sensitivity of a system's stationary states to slow driving. We probe this sensitivity via the system's susceptibility to the average protocol speed, which we call the ``speed-Fisher information," and relate it to irreversible entropy production in the system. We show that chaotic dynamics manifests as a divergence of the speed-Fisher information with the protocol time, and that this response is controlled by the perturbation's low-frequency spectral weight. This approach to chaos applies to both classical and quantum Hamiltonian systems, and naturally extends to non-Hamiltonian classical flows. We illustrate this framework with simple classical and quantum examples, along with a non-Hamiltonian flow that qualitatively exhibits analogous low-frequency spectral behavior.
\end{abstract}

\maketitle

Chaos is traditionally understood through the presence of long-time unpredictability in a system, despite the system's dynamics being governed by deterministic laws~\cite{ott_1993}. \refD{While classical chaos is diagnosed through the exponential instabilities of a system's trajectories~\cite{lorenz_1972,gleick_1987,oseledets_1968}, quantum systems do not admit such a description since unitary evolution preserves distinguishability between any two quantum states~\cite{berry_1989}. In systems with a finite number of degrees of freedom, quantum chaos is commonly diagnosed through spectral signatures~\cite{bohigas_1984,berryTabor_1977} such as level repulsion and Wigner-Dyson statistics~\cite{wigner_1951aa,dyson_1962aa,dyson_1962ab,dyson_1962ac}.} Because these particular metrics for quantum and classical chaos are fundamentally distinct, they do not represent a comprehensive framework that seamlessly explains (i) the emergence of classical chaos from the underlying quantum dynamics and (ii) the emergence of long-term dynamical instabilities in quantum systems, especially close to the integrable, non-chaotic limits.

\refD{Recent work has established the adiabatic gauge potential (AGP)~\cite{kolodrubetz_2017}, the generator of adiabatic deformations, as a sensitive probe of both classical and quantum chaos~\cite{pandey_2020,kim_2026,karve_2025aa}. Building on these results, we propose that the sensitivity of stationary states to finite-speed driving provides an operational probe of chaos.} \refA{The adiabatic theorem states that sufficiently slow Hamiltonian deformations preserve adiabatic invariants in classical systems and produce adiabatic following of instantaneous eigenstates in quantum systems~\cite{dirac_1925,arnold_1989,messiah_2014}. Because such adiabatic behavior can occur in both regular and chaotic systems, the strict adiabatic limit does not by itself distinguish between them. Their leading response at finite driving speed, however, can differ markedly. We therefore characterize chaos through the susceptibility of a stationary state to the driving speed of an applied perturbation.}

\begin{figure}
    \centering
    \includegraphics[scale=1]{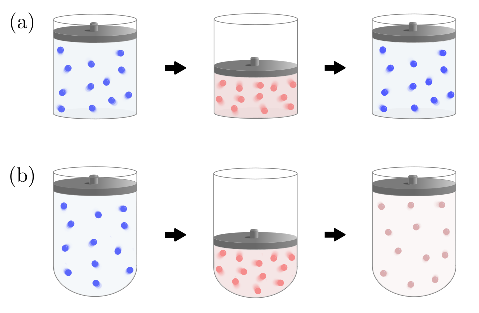}
    \caption{\refA{Slow, cyclic driving of (a) a regular system, whose final state closely approaches its initial state, and (b) a chaotic system, where thermodynamic friction produces dissipated work at finite driving speeds.}}
    \label{fig_sketch}
\end{figure}

\refA{Our construction considers stationary states subjected to slow cyclic driving. As we show in this Letter, chaotic dynamics can sustain a nonzero, or even a divergent, thermodynamic friction coefficient in the adiabatic limit (see Fig.~\ref{fig_sketch}).} We call the corresponding susceptibility to the speed of the drive the ``speed-Fisher information"~\cite{fisher_1922}. We show that this Fisher information is governed by the low-frequency spectral weight of physical observables and can be interpreted as a thermodynamic drag, making it accessible both numerically and experimentally. This framework applies uniformly to quantum and classical systems, including non-Hamiltonian deterministic flows, as we demonstrate through representative examples. \refD{While we present a self-contained discussion of our results below, the full technical details can be found in the companion paper~\cite{karve_companion}.}

Consider a system described by the Hamiltonian $H_0$, which is prepared in a stationary state $\rho_{-\infty}$ at time $\tau=-\infty$. Depending on the context, $\rho_{-\infty}$ corresponds either to a probability distribution in the classical phase-space or to a quantum density matrix. The arguments that follow in this Letter apply to both classical and quantum systems, and therefore, all expressions are agnostic as to whether $\rho_{-\infty}$ is interpreted as a probability distribution or a density matrix. In the absence of additional conserved quantities, it is natural to take this initial stationary state to be a function of the Hamiltonian alone; that is,
\begin{equation}
    \rho_{-\infty} = P(H_0).
\end{equation}
\refD{Although integrable systems admit stationary ensembles depending on additional conserved quantities, we restrict attention solely to states that have the above form, with a smooth $P$. Our conclusions about the scaling of the speed-Fisher information do not depend on this assumption.}

\refA{We probe the stability of this state using a cyclic perturbation $H(t) = H_0 + \lambda(\tau)V$, with $\lambda(\pm\infty) = 0$. The perturbation $\lambda(\tau)$ depends on two key parameters: the maximum amplitude $\lambda_{\text{max}}$ and the duration of the protocol $t$. Since we are interested in the susceptibility of the final state to the driving speed, it is convenient to trade the amplitude for the mean speed of the protocol. We therefore parameterize the perturbation as
\begin{equation}
    \lambda(\tau) = \frac{\bar{v}}{\mu} f(\mu \tau),
    \label{eq_parameterization}
\end{equation}
where $\mu = \nicefrac{1}{t}$ is the inverse duration and $\bar{v}$ is the average speed, defined as
\begin{equation}
    \bar{v} = \frac{1}{t} \int_{-\infty}^\infty d\tau \ \left|\dot{\lambda}(\tau)\right|.
\end{equation}
Additionally, $f(x)$ is a smooth function, satisfying $f(\pm\infty)=0$, along with the normalization $\int_{-\infty}^{\infty} |g(x)| \ dx = 1$, where $g(x) = f'(x)$~\footnote{Naively, it may appear that the perturbation strength in Eq.~(\ref{eq_parameterization}) diverges as $\mu\to 0$, barring us from applying perturbation theory. However, note that the speed-Fisher information is defined in the limit $\bar{v}\to 0$ as $\mu$ is held fixed, ensuring that the perturbative result is valid at any $\mu$. In experiments where one must operate at some nonzero speed, it must be ensured that $\bar{v} \ll \mu$ to keep the system in the perturbative regime. Moreover, this issue becomes irrelevant in the cases where the perturbation $\lambda(\tau)$ corresponds to an angle variable, since such an angle can take any arbitrarily large value.}. Moreover, we assume that $\tilde{g}(\omega)$ -- the Fourier transform of $g(x)$ -- has a sufficiently fast decaying high-frequency tail, such that $|\tilde{g}(\omega)| \leq \mathcal{O}(\omega^{-1})$. Additionally, $\tilde{g}(0) = 0$ since the protocol is cyclic.
For a fixed protocol shape, $(\lambda_{\text{max}}, t)$ and $(\bar{v},\mu)$ are equivalent parameterizations.}

We measure the overlap between the initial and final states $\rho_{\pm\infty}$ through the fidelity~\cite{jozsa_1994,nielsen_2010}, which depends on both the speed and duration of the protocol, and thus, we denote it as $\mathscr{F}(\bar{v},\mu)$. This fidelity can be computed classically through the Hellinger distance~\cite{hellinger_1909} or quantum mechanically through the Bures distance~\cite{bures_1969,helstrom_1967}. The fidelity obeys $\mathscr{F}(\bar{v},\mu) \leq 1$, with equality if and only if the final state coincides with the initial state.

\begin{table}
\begin{ruledtabular}
    \vspace*{1em}
    \begin{tabular}{lll}
        & $\tilde{\mathcal{D}}_{V}(\omega\to 0)$ & $\mathscr{I}_{\bar{v}}(\mu\to 0)$ \\
        \midrule
        Regular & $\sim |\omega|^{\alpha} \ (\alpha > 1)$ & $\mathcal O\left(\mu^0\right)$ \\
        \midrule
        \makecell[l]{Marginally \\ unstable} & $\sim |\omega|^{\alpha}\ (0<\alpha \leq 1)$ & $\mathcal{O}\left({\nicefrac{1}{\mu^{1\!-\!\alpha}}}\right)$ \\
        \midrule
        \makecell[l]{Chaotic \\ thermalizing} & $\text{const} > 0$ & $\mathcal{O}\left({\nicefrac{1}{\mu}}\right)$ \\
        \midrule
        \makecell[l]{Chaotic \\ nonthermalizing} & $\sim |\omega|^{\alpha} \ (\alpha < 0)$ & $\mathcal{O}\left({\nicefrac{1}{\mu^{1\!+\!|\alpha|}}}\right)$ \\
    \end{tabular}
\end{ruledtabular}
\caption{Relation between the low-frequency spectral weight and the speed-Fisher information.}
\label{tab_chaos}    
\end{table}

The fidelity approaches $1$ as $\bar{v} \rightarrow 0$. Small but nonzero drive speeds leave an imprint on the system's final state. We capture the system's susceptibility to drive speed by defining the ``speed-Fisher information" as
\begin{equation}
    \mathscr{I}_{\bar{v}}(\mu) = -2\frac{\partial^2\mathscr{F}(\bar{v},\mu)}{\partial \bar{v}^2} \Biggr|_{\bar{v}=0}.
\end{equation}
\refA{This susceptibility is evaluated at a fixed value of $\mu$ while taking the limit $\bar{v}\to 0$.} Equivalently,
\begin{equation}
    \mathscr{F}(\bar{v},\mu) = 1 - \frac{1}{4}\mathscr{I}_{\bar{v}}(\mu)\bar{v}^2 + \mathcal{O}(\bar{v}^3).
\end{equation}
\refA{Note the absence of a linear correction to the fidelity, due to both the initial and final states being normalized.} Thus, $\mathscr{I}_{\bar{v}}(\mu)$ in the limit $\mu\to 0$ captures the leading loss of fidelity due to a nonzero drive speed of a quasistatic process. \refA{The larger $\mathscr{I}_{\bar{v}}(\mu)$ is, the more susceptible the state is to changes in the drive's speed. We therefore use the quasistatic scaling of $\mathscr I_{\bar v}(\mu)$ as an operational probe of chaos. As shown below, generic perturbations of chaotic systems produce substantial low-frequency spectral weight, causing $\mathscr I_{\bar v}(\mu)$ to diverge as $\mu\to 0$ \footnote{It should be noted that this statement is valid as long as $\nicefrac{1}{\mu}$ is much smaller than the Heisenberg time in quantum systems.}.}

The relevant equilibrium object that determines this Fisher information is not the bare correlation function of $V$, but a score-weighted symmetric autocorrelation function, defined as
\begin{equation}
    \mathcal{D}_V(\tau) = \frac{1}{2}\left\langle s(H_0)^2 \ [\delta V(\tau) , \delta V(0)]_+ \right\rangle,
\end{equation}
where $[ \ , \ ]_+$ denotes the anticommutator, $\langle \ \rangle$ denotes an average over $\rho_{-\infty}$, and $\delta V(\tau) = V(\tau) - \bar{V}$ is the perturbation with its infinite time-average removed, where $V(\tau)$ is evolved under the unperturbed Hamiltonian $H_0$. The score is defined as
\begin{equation}
    s(H_0) = \frac{\partial \ln \rho_{-\infty}}{\partial E}\bigg|_{E=H_0}\equiv  \frac{\partial \ln P(E)}{\partial E}\bigg|_{E=H_0}\,.
\end{equation}
\refB{On the other hand, the classical score-weighted autocorrelation is given by
\begin{equation}
    \mathcal{D}_V(\tau) = \left\langle s(H_0)^2  \delta V(\tau) \delta V(0) \right\rangle,
\end{equation}
where $V(\tau)$ is again evolved under the unperturbed Hamiltonian $H_0$. For smooth observables and stationary distributions, the quantum expression reduces to its classical counterpart in the semiclassical limit $\hbar\to 0$: in the Wigner–Weyl representation, the symmetrized operator product becomes the classical product, and the quantum expectation becomes a phase-space average~\cite{polkovnikov_2010}.}

Notably, for the most common Gibbs ensemble with $\rho_{-\infty} = \frac{1}{Z}e^{-H_0/\kB T}$, the score is simply $\nicefrac{-1}{\kB T}$, so that the score-weighted correlation function is proportional to the standard one:
\begin{equation}
    \mathcal{D}_V(\tau) = \frac{(\kB T)^{-2}}{2}\left\langle \ [\delta V(\tau) , \delta V(0)]_+ \right\rangle.
\end{equation}

As shown in~\cite{karve_companion}, the speed-Fisher information can be written as
\begin{equation}
\label{eq:FI_mu}
    \mathscr{I}_{\bar{v}}(\mu) = \frac{1}{\mu}\int_{-\infty}^\infty \frac{d\omega}{2\pi} \ \tilde{\mathcal{D}}_V(\mu \omega) \left|\tilde{g}(\omega)\right|^2,
\end{equation}
where the score-weighted spectral function of $V$: $\tilde{\mathcal{D}}_V(\omega)$, is defined as the Fourier transform of $\mathcal{D}_V(\tau)$. \refB{This equation connects the speed-Fisher information to AGP-based diagnostics of chaos: both probe the low-frequency spectral weight of the perturbation. In fact, for the Gibbs state, the Fisher information is identical, up to a factor, to the regularized AGP norm for the protocol $\lambda(\tau) = \frac{\bar{v}}{2\mu}e^{-\mu|\tau|}$ with the Lorentzian filter function $\tilde g(\omega)\propto \tfrac{\omega}{1+\omega^2}$ (cf. Eq.~\eqref{eq:FI_mu} with Eq. (25) in Ref.~\cite{kim_2026}). However, in general, the speed-Fisher information depends on the specific cyclic protocol acting on a chosen stationary state}. 

In the above expression, the shape of the protocol enters only through the filter function 
$|\tilde g(\omega)|^2$. The quasistatic scaling is therefore controlled by a combination of the low-frequency behavior of the score-weighted spectral function and the high-frequency tail of the protocol filter. For regular or integrable dynamics, the spectral functions of observables corresponding to integrability-preserving perturbations or weak integrability-breaking perturbations \cite{surace_2023,vanovac_2024,vanovac_2026} vanish at low frequency, at least linearly and typically quadratically. This condition guarantees perturbative stability of classical orbits/quantum eigenstates~\cite{kim_2026}. As a result, the speed-Fisher information does not diverge in the quasistatic limit: depending on the protocol shape, it either vanishes with $\mu$ or saturates to a finite value. By contrast, generic perturbations in chaotic thermalizing systems, i.e., systems which satisfy the eigenstate thermalization hypothesis (ETH)~\cite{deutsch_1991,srednicki_1994,dalessio_2016}, have nonzero low-frequency spectral weight, producing a robust $\nicefrac{1}{\mu}$ divergence of $\mathscr I_{\bar v}$, \refA{in agreement with Ott's diffusive scaling of adiabatic invariants in classical systems~\cite{ott_1979}}. Nonthermalizing chaotic systems can have singular low-frequency spectra, see Refs.~\cite{leblond_2020,kim_2026}, leading to an even stronger divergence. Finally, spectral functions with $0 < \alpha < 1$ are found in critical~\cite{pereira_2008}, or sub-Ohmic systems~\cite{leggett_1987}, where the speed-Fisher information exhibits a weaker divergence. See Table~\ref{tab_chaos} for a summary of the different regimes. \refB{This spectral characterization probes long-time correlations of physical observables, complementing Lyapunov exponents, which instead quantify the instability of nearby classical trajectories. Thus, timescales probed by spectral diagnostics can be different from Lyapunov timescales~\cite{kim_2025}. Where these correlations admit a Ruelle–Pollicott resonance description~\cite{ruelle_1986}, the resonances contribute to $\tilde{\mathcal D}_V(\omega)$ and hence to the speed-Fisher information, without needing to be resolved individually.}

The speed-Fisher information has a direct thermodynamic interpretation: as a friction that leads to irreversible ``modular energy" production under the cyclic perturbation. To see this, consider the modular Hamiltonian of the initial stationary state, defined as
\begin{equation}
    K_0 = -\ln \rho_{-\infty}.
\end{equation} 
To leading order in the perturbation, one finds that
\begin{equation}
    D_{\text{KL}}(\rho_{+\infty}||\rho_{-\infty}) = \langle\Delta K_0\rangle = \frac{1}{2}\mathscr{I}_{\bar{v}}(\mu)\bar{v}^2,
    \label{eq_thermDrag}
\end{equation}
where $D_{\text{KL}}(\rho_{+\infty}||\rho_{-\infty})$ is the relative entropy~\cite{kullback_1951} between the final and initial states. See~\cite{karve_companion} for a detailed derivation. In particular, when $\rho_{-\infty}$ is a Gibbs state, the modular Hamiltonian is simply $K_0 = \frac{1}{\kB T} H_0 + \ln Z$. Thus, the Fisher information is the drag~\cite{dalessio_2013} associated with irreversible heating, since
\begin{equation}
    \langle\Delta E\rangle = \frac{\kB T}{2}\mathscr{I}_{\bar{v}}(\mu)\bar{v}^2.
    \label{eq_thermDragGibbs}
\end{equation}
\refD{This implies that the thermodynamic friction coefficient is proportional to $\mu \mathscr{I}_{\bar{v}}(\mu)$~\cite{dalessio_2013}. Note that the friction coefficient vanishes in regular systems, remains finite in chaotic thermalizing systems, and diverges in nonthermalizing systems.} Thus, measuring the change in the modular energy provides us with a direct way to measure the speed-Fisher information in experimental or numerical setups. \refB{This provides an operational route to the speed-Fisher information through measurements of the dissipated work, without reconstructing the final state or explicitly determining the AGP.} Alternatively, one can express the Fisher information in terms of the fluctuations of $\Delta K_0$ by using the generalized Jarzynski identity~\cite{jarzynski_1997}
\begin{equation}
    \left\langle e^{-\Delta K_0} \right\rangle = 1.
\end{equation}
Expanding this identity in cumulants, one finds that
\begin{equation}
    \text{Var}(\Delta K_0)\approx {\bar{v}^2} \mathscr{I}_{\bar{v}}(\mu) ,
    \label{eq_Evar}
\end{equation}
assuming that other higher-order cumulants of $\Delta K_0$ are negligible. 

\refD{Having established the relation between susceptibility to the speed of a drive and the perturbation's low-frequency spectral weight, we turn to some representative examples. Throughout, we employ the cyclic protocol
\begin{equation}
    \lambda(\tau) =
    \begin{cases}
        \dfrac{\bar{v}}{4\mu}\sin^2(2\pi\mu\tau), & 0\leq \tau\leq \dfrac{1}{\mu},\\
        0, & \text{otherwise},
    \end{cases}
    \label{eq_drive}
\end{equation}
which has duration $\nicefrac{1}{\mu}$, average speed $\bar v$, and maximum amplitude $\lambda_{\text{max}}=\nicefrac{\bar v}{4\mu}$. \refB{The corresponding filter function is
\begin{equation}
\tilde{g}(\omega)
= \frac{2\pi^2(1-e^{-i\omega})}{16\pi^2-\omega^2},
\label{eq_filterFn}
\end{equation}
which satisfies the condition $|\tilde g(\omega)|\leq \mathcal{O}(|\omega|^{-1})$ at high frequencies.} For each $\mu$, we evaluate the modular-energy change at several values of $\lambda_{\text{max}}$ and extract the speed-Fisher information from either $\nicefrac{2\langle\Delta K_0\rangle}{\bar v^2}$ or $\nicefrac{\text{Var}(\Delta K_0)}{\bar v^2}$. Independence of these quantities from $\lambda_{\text{max}}$ identifies the perturbative regime and yields the $\bar v\to0$ limit. Unless stated otherwise, we estimate this limit by averaging over $\lambda_{\text{max}}\in[0.1,1]$. Repeating this procedure as $\mu$ is decreased determines the quasistatic scaling of $\mathscr{I}_{\bar v}(\mu)$.
}

\begin{figure*}[htpb]
    \centering
    \begin{subfigure}[t]{0.32\linewidth}
        \includegraphics[width=\linewidth]{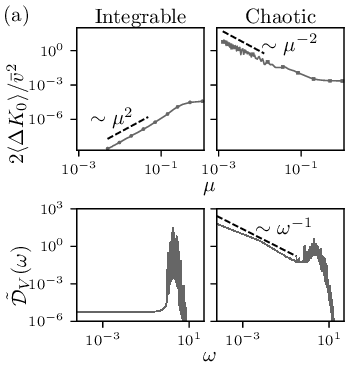}
        \phantomcaption
        \label{fig_2spinFI}
    \end{subfigure}
    \hfill
    \begin{subfigure}[t]{0.32\linewidth}
        \includegraphics[width=\linewidth]{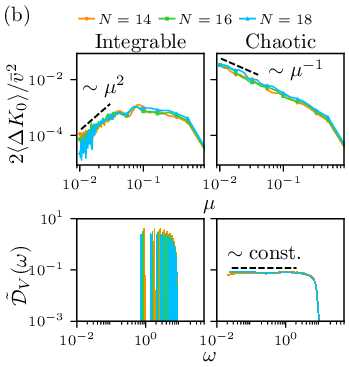}
        \phantomcaption
        \label{fig_isingFI}
    \end{subfigure}
    \hfill
    \begin{subfigure}[t]{0.32\linewidth}
        \includegraphics[width=\linewidth]{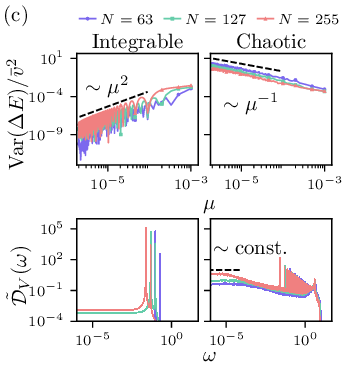}
        \phantomcaption
        \label{fig_fputFI}
    \end{subfigure}
    \caption{\raggedright \refC{Estimates of the speed-Fisher information as a function of $\mu$ (top panels) and the corresponding normalized equilibrium spectral function as a function of $\omega$ (bottom panels) for three different examples. \refA{In all cases, the Fisher information is averaged over $\lambda_{\text{max}}\in [0.1,1]$ to obtain the perturbative limit $\bar{v}\to 0$.} (a) The integrable (left) and chaotic (right) quantum two-spin models, with $S=10$. Parameters used: integrable model -- $\mathbf{J} = (1,1,1/2)$, $\mathbf{A} = (0,0,0)$, $\kB T=100$, chaotic model -- $\mathbf{J} = (3/2,\pi,\sqrt{e})$, $\mathbf{A} = (\sqrt{\pi},\sqrt{3},e)$, $\kB T=100$. (b) The integrable (left) and chaotic (right) Ising models. Parameters used -- integrable model: $J = 1$, $h_x=1.05$, $h_z=0$, $\kB T=100$. chaotic model: $J = 1$, $h_x=1.05$, $h_z=0.5$, $\kB T=100$. (c) The classical $\beta$-FPUT model with $\beta=0$ (left) and $\beta=10$ (right), initialized in a microcanonical ensemble with energy $E=1$.} }
\end{figure*}

\refC{Let us first consider a semi-classical few-body system composed of two spins, described by the following Hamiltonian:
\begin{equation}
    H_0 = \sum_{\alpha \in\{x,y,z\}} \left[-J_{\alpha} S_{1\alpha} S_{2\alpha} + \frac{1}{2} A_\alpha \left(S_{1\alpha}^2 + S_{2\alpha}^2\right)\right],
\end{equation}
where $\mathbf{S}_1 = (S_{1x},S_{1y},S_{1z})$ and $\mathbf{S}_2 = (S_{2x},S_{2y},S_{2z})$ are spin-$S$ degrees of freedom in the quantum version, and are unit vectors in the classical case. Depending on the choice of parameters $\mathbf{J} = (J_x,J_y,J_z)$ and $\mathbf{A}=(A_x,A_y,A_z)$, this model can be integrable or chaotic~\cite{kim_2026}. In either case, we drive this system through the perturbation $V = S_{1z}S_{2z}$.

We sample states from a Gibbs ensemble and measure the energy absorbed by the system due to the drive, from which the thermodynamic drag is extracted as per Eq.~(\ref{eq_thermDragGibbs}). This is plotted as a function of $\mu$ in the upper panel of Fig. \ref{fig_2spinFI} for both the integrable and chaotic versions. As expected, the speed-Fisher information vanishes in the slow-driving limit of the integrable system, while it diverges approximately as $\nicefrac{1}{\mu^2}$ in the chaotic case. Consistent with this behavior, we plot the corresponding classical spectral function of the observable $S_{1z}S_{2z}$ in the bottom panel of Fig. \ref{fig_2spinFI}. While the spectral function of the integrable model decays to a numerical floor as $\omega\to 0$, it diverges as $\approx \nicefrac{1}{\omega}$ in the chaotic case for the parameters analyzed here. The extracted exponents for the spectral divergence and the Fisher information growth are consistent with our predicted scaling relation and suggest that this system belongs to the chaotic, nonthermalizing category.

Next, we turn our attention to a paradigmatic many-body quantum system: the one-dimensional, spin-$1/2$ Ising model with both transverse and longitudinal fields~\cite{ovchinnikov_2003}, described by the Hamiltonian:
\begin{equation}
    H = J \sum_{i=1}^N \sigma^z_i \sigma^z_{i+1} + h_x \sum_{i=1}^N \sigma^x_i + h_z \sum_{i=1}^N \sigma^z_i,
\end{equation}
with periodic boundary conditions ($\mathbf{\sigma}_1 = \mathbf{\sigma}_{N+1}$). In the absence of the longitudinal field ($h_z=0$), this model reduces to the transverse-field Ising chain, which is integrable~\cite{pfeuty_1970}. A longitudinal field generically breaks integrability, and the resulting mixed-field Ising model exhibits chaotic behavior~\cite{banuls_2011}. In either case, we drive this system through the perturbation $V = \displaystyle\sum_{i=1}^N \sigma^x_i$.

The speed-Fisher information is plotted as a function of $\mu$ in the top panel of Fig. \ref{fig_isingFI}. \refA{In the integrable case, the Fisher information decreases toward zero as $\mu \to 0$, indicating vanishing thermodynamic drag. By contrast, in the chaotic thermalizing (ETH) regime, the Fisher information diverges as $\mathscr{I}_{\bar v}(\mu) \sim \nicefrac{1}{\mu}$, signaling the presence of a finite friction coefficient in the quasistatic limit.} This behavior is consistent with the spectral function of $V$, plotted in the bottom panel of Fig. \ref{fig_isingFI}. While the spectral function of the integrable model exhibits discrete peaks and lacks low-frequency weight, the chaotic case exhibits a low-frequency plateau. Thus, numerically, this shows that the mixed-field Ising model belongs to the chaotic, thermalizing category.

Finally, we consider the classical $\beta$-Fermi-Pasta-Ulam-Tsingou ($\beta$-FPUT) model~\cite{fermi}, which is a one-dimensional chain of oscillators described by the following Hamiltonian:
\begin{equation}
    H_0 = \sum_{n=1}^{N} \frac{p_n^2}{2} + \frac{1}{2}\sum_{n=0}^N (q_{n+1}-q_n)^2 + \frac{\beta}{4}\sum_{n=0}^N (q_{n+1}-q_n)^4,
\end{equation}
subject to fixed boundary conditions, $q_0=q_{N+1}=0$. For $\beta=0$, the model reduces to a harmonic chain and is integrable. For a nonzero $\beta$, the quartic interaction breaks integrability. This model is known to exhibit metastability at small $\beta$, but thermalizes quickly at larger nonlinearities~\cite{benettin_2008,reiss_2023}. Here, we consider models with large numbers of particles, and apply a quartic perturbation $\displaystyle\frac{\lambda(\tau)}{4} \sum_{n=0}^N (q_{n+1}-q_n)^4$, where $\lambda(\tau)$ is given by Eq.~(\ref{eq_drive}). We initialize this system in a microcanonical ensemble with energy $E=1$, which is, due to the large number of particles considered here, locally equivalent to a Gibbs state with temperature $\kB T = \left(\frac{\partial S}{\partial E}\right)^{-1}$~\cite{touchette_2015}. Then, Eq.~(\ref{eq_Evar}) allows us to extract the scaling of the speed-Fisher information with $\mu$ from $\nicefrac{\text{Var}(\Delta E)}{\bar{v}^2}$.

In Fig. \ref{fig_fputFI}, we consider two extremes of the $\beta$-FPUT model: the integrable case with $\beta=0$, and a strongly nonlinear case with $\beta=10$. As expected, the susceptibility vanishes in the integrable case as $\mu\to 0$, and it diverges as $\nicefrac{1}{\mu}$ when $\beta=10$. This is consistent with the behavior of the corresponding spectral function at small frequencies: it vanishes in the integrable regime and approaches a finite low-frequency plateau in the chaotic (ETH) regime.
}

While all our above examples are restricted to Hamiltonian systems, we also discuss how this framework extends to non-Hamiltonian systems in the End Matter. For classical autonomous systems, a perturbation corresponds to a velocity field, and the relevant observable in such a case is the divergence of the perturbation. As discussed therein, the stability of a stationary state under slow, cyclic driving is again related to the low-frequency spectral weight of this divergence, and analogous low-frequency spectral behavior of generic observables is demonstrated in the Lorenz-96 system~\cite{lorenz_1996}.

In summary, in this Letter, we discuss an operational notion of chaos that treats classical and quantum systems on an equal footing. Instead of studying phase-space trajectories or level-spacing statistics of the system, we argue that chaos is linked to the response of a stationary state under slow driving. We introduce the speed-Fisher information, which quantifies the susceptibility of the stationary state to the average speed of the drive, and show that it can be interpreted as a thermodynamic drag. This Fisher information is also directly related to entropy generation and energy absorption. The robustness of this framework is demonstrated by probing chaos via this drag in a semi-classical two-spin model, a quantum mixed-field Ising chain, and the classical $\beta$-FPUT model, as well as the Lorenz-96 system in the End Matter.

\refC{In contrast to traditional perspectives, our framework posits that regular or chaotic behavior is a property not only of the system and its state, but also of the relevant observable, as reflected in its low-frequency spectral behavior. In chaotic thermalizing systems, most local observables are expected to reach equilibrium at the same time scale, known as the Thouless time~\cite{schiulaz_2019,suntajs_2020,suntajs_2022}. However, certain observables, such as currents, can relax faster, leading to a more stable response to perturbations coupled to them. Additionally, conserved perturbations (i.e. $[H_0,V]=0$) may completely fail to reveal the chaotic nature of the system, since they leave the stationary state unchanged. The difference between different observables becomes more pronounced close to integrable regimes~\cite{kim_2025}, as is often observed in real dynamical systems~\cite{mogavero_2023}. This reflects the fact that some observables, such as the total energy in a closed system, are not sensitive to whether the dynamics is regular or chaotic. We also highlight that in large systems, whether quantum or classical, microscopic degrees of freedom are usually inaccessible. So any experimentally relevant probe of chaos must involve observables. Our framework is therefore complementary to traditional diagnostics, offering not only a criterion for the presence of chaos, but a more detailed description of how chaos manifests across physical observables and timescales.}

\textit{\textbf{Acknowledgments}} -- The authors thank Hyeongjin Kim, Guilherme Delfino, and Bernardo Barrera for insightful discussions. AP was supported by: NSF grant no. DMR-2412542 and AFOSR grant no. FA9550-21-1-0342. The authors acknowledge the use of Boston University’s Shared Computing Cluster (SCC) for numerical simulations.

\textit{\textbf{Data Availability Statement}} -- The data that support the findings of this study were generated by numerical simulations. The data, source code, and parameters used to generate the simulations are publicly available~\cite{karve_ising,karve_2026_beta_fput,karve_2026_dataset_2spin,karve_2026_dataset_lorenz}.

\bibliography{references}

\begin{figure*}[htb]
    \centering
    \begin{subfigure}[t]{0.47\textwidth}
        \centering
        \includegraphics[width=\linewidth]{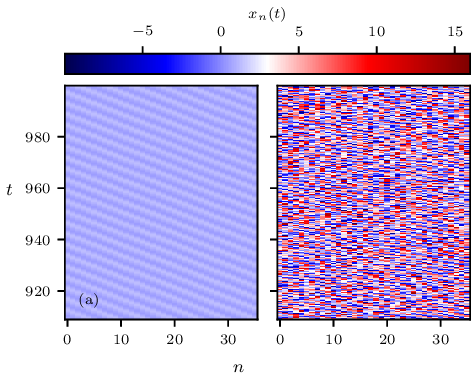}
        \phantomcaption
        \label{fig_lorenz96_a}
    \end{subfigure}
    \hfill
    \begin{subfigure}[t]{0.47\textwidth}
        \centering
        \includegraphics[width=\linewidth]{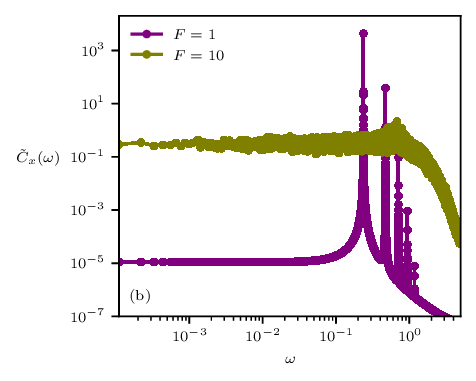}
        \phantomcaption
        \label{fig_lorenz96_b}
    \end{subfigure}
    \caption{(a) Time evolution of a $N=36$ site Lorenz-96 system with $F=1$ (left) and $F=10$ (right). (b) Spectral function of $x$.}
\end{figure*}

\textit{\textbf{Appendix: Non-Hamiltonian Systems}} -- Remarkably, the framework developed in this Letter also extends to non-Hamiltonian, autonomous systems, such as those described by a differential equation of the form
\begin{equation}
\label{eq:map_pert}
    \frac{d\mathbf{x}}{d\tau} = \mathbf{F}(\mathbf{x}) + \bar{v} f(\mu \tau)\mathbf{V}(\mathbf{x}),
\end{equation}
where the state of the system is described by a multi-dimensional vector $\mathbf{x}$. Note that since the perturbation $\mathbf{V}$ is itself a velocity field, the parameter $\bar{v}$ describes the average drive speed. Alternatively, we can understand that the Hamiltonian evolution in the instantaneous basis is given by the moving frame Hamiltonian $H_\text{mv}=H-\dot \lambda \mathcal A_\lambda$, where $\mathcal A_\lambda$ is the adiabatic gauge potential~\cite{kolodrubetz_2017}. The second boost term is responsible for all non-adiabatic effects and plays a similar role as the last term in Eq.~\eqref{eq:map_pert}. In such autonomous systems, the dressed autocorrelation is defined as $\mathcal{D}_V(\tau) = \langle L_V(\tau)L_V(0) \rangle_c$, where
\begin{equation}
    L_V(\mathbf{x}) = \frac{1}{\rho(\mathbf{x},-\infty)}\mathbf{\nabla}\cdot(\rho(\mathbf{x},-\infty)\mathbf{V}(\mathbf{x})),
\end{equation}
and $\rho(\mathbf{x},-\infty)$ is the initial stationary density. The Fisher information is then given by
\begin{equation}
    \mathscr{I}_{\bar{v}}(\mu) = \frac{1}{\mu}\int_{-\infty}^\infty \frac{d\omega}{2\pi} \ \tilde{\mathcal{D}}_V(\mu \omega) \left|\tilde{f}(\omega)\right|^2.
\end{equation}

\refB{In non-Hamiltonian autonomous systems, the Gibbs entropy of the system, $S$, itself need not be conserved, where
\begin{equation}
    S = -\int d\mathbf{x} \ \rho(\mathbf{x})\ln\rho(\mathbf{x}).
\end{equation}
Thus, a more general version of Eq.~(\ref{eq_thermDrag}) is
\begin{equation}
    \langle \Delta K_0 \rangle - \Delta S \simeq \frac{1}{2}\mathscr{I}_{\bar{v}}(\mu)\bar{v}^2,
\end{equation}
with the modular Hamiltonian given by
\begin{equation}
    K_0(\mathbf{x}) = -\ln\rho_{-\infty}(\mathbf{x}).
\end{equation}
Volume-preserving dynamics guarantees $\Delta S=0$, whereas compressible flows generally produce an additional entropy contribution.

Note that the above arguments require that the initial stationary state be smooth. However, such systems can admit non-smooth stationary states, rendering $L_V$ and $K_0$ ill-defined.} Nevertheless, one can infer the chaotic nature of such systems from the low-frequency spectral behavior of generic observables. As an example, consider the Lorenz-96 system~\cite{lorenz_1996}, defined as
\begin{equation}
    \frac{dx_n}{d\tau} = (x_{n+1}-x_{n-2}) x_{n-1} - x_n + F,
\end{equation}
for $n=0,1,\dots,N-1$, and with periodic boundary conditions. The Lorenz--96 model is a standard low-dimensional toy model for atmospheric dynamics that captures advection, dissipation, and an external forcing term $F$~\cite{karimi_2010}. For small $F$, the dynamics is regular, while for sufficiently large $F$, the system becomes chaotic, as shown in Fig. \ref{fig_lorenz96_a}. This behavior is also reflected in the spectral function of a generic observable of this model. As shown in Fig. \ref{fig_lorenz96_b}, the spectral function of a generic observable, $x$, has a low-frequency noise floor when $F$ is small. On the other hand, in the chaotic system, the spectral function of the same observable develops a low-frequency plateau.

\end{document}